# Integrated electro-optics on thin-film lithium niobate

Yaowen Hu[1,2,*], Di Zhu[3,4], Shengyuan Lu[1], Xinrui Zhu[1], Yunxiang Song[1], Dylan Renaud[1], Daniel Assumpcao[1], Rebecca Cheng[1], CJ Xin[1], Matthew Yeh[1], Hana Warner[1], Xiangwen Guo[5], Amirhassan Shams-Ansari[1,6], David Barton[1,7], Neil Sinclair[1], and Marko Loncar[1,*]

*[1]John A. Paulson School of Engineering and Applied Sciences, Harvard University, Cambridge, MA 02138, USA*
*[2]State Key Laboratory for Mesoscopic Physics and Frontiers Science Center for Nano-optoelectronics, School of Physics, Peking University, Beijing 100871, China*
*[3]Department of Materials Science and Engineering, National University of Singapore, Singapore 117575, Singapore*
*[4]Institute of Materials Research and Engineering (IMRE), Agency for Science, Technology and Research (A*STAR), Singapore 138634, Singapore*
*[5]Department of Electrical and Computing Engineering, University of Virginia, Charlottesville, Virginia 22903, USA*
*[6]DRS Daylight Solutions, 16465 Via Esprillo, San Diego, CA, USA*
*[7]Department of Materials Science and Engineering, Northwestern University, Evanston, Illinois 60208, USA*
*[*]Corresponding authors: yaowenhu@pku.edu.cn; loncar@seas.harvard.edu;*

**Electro-optics serves as the crucial bridge between electronics and photonics, unlocking a wide array of applications ranging from communications and computing to sensing and quantum information. Integrated electro-optics approaches in particular enable essential electronic high-speed control for photonics while offering substantial photonic parallelism for electronics. Recent strides in thin-film lithium niobate photonics have ushered revolutionary advancements in electro-optics. This technology not only offers the requisite strong electro-optic coupling but also boasts ultra-low optical loss and high microwave bandwidth. Further, its tight confinement and compatibility with nanofabrication allow for unprecedented reconfigurability and scalability, facilitating the creation of novel and intricate devices and systems that were once deemed nearly impossible in bulk systems. Building upon this platform, the field has witnessed the emergence of various groundbreaking electro-optic devices[1–12] surpassing the current state of the art[1–6,9–12], and introducing functionalities that were previously non-existent[3,7,8]. This technological leap forward provides a unique framework to explore various realms of physics as well, including photonic non-Hermitian synthetic dimensions[13–15], active topological physics[16,17], and quantum electro-optics[12,18–20]. In this review, we present the fundamental principles of electro-optics, drawing connections between fundamental science and the forefront of technology. We discuss the accomplishments and future prospects of integrated electro-optics, enabled by thin-film lithium niobate platform.**

## Introduction

Over the past century, photonic and electronic systems, both governed by the laws of electromagnetism, have become indispensable in advancing science and technology. The evolution of modern electronic systems, culminating in the creation of integrated circuits, has revolutionized our daily lives, influencing everything from computers and cell phones to automated control systems and robots. Simultaneously, integrated photonic systems offering broad bandwidth, minimal propagation loss, and massive parallelism, enabled by breakthroughs in micro- and nano-fabrication technologies have reshaped fields such as sensing, imaging, metrology, spectroscopy, and bio-medicine. At the intersection of these two dynamic fields is the field of electro-optics that harnesses distinct yet complementary properties of photons and electrons. On one end, electro-optics endows passive photonic systems with fast control provided by high-speed electronics, thus enabling modern communications[21–25], lidar[26,27], and optical coherent tomography[28,29] systems. Conversely, photonics enriches electronics with the massively parallel processing capabilities, evident in advancements in photonic computing[30–34] and microwave photonics[35–39], for example. These application-driven developments bridge the electronic and optical domains, leveraging materials that offer control of photons using electrical signals. While several physical mechanisms can be employed to achieve this, including thermo-optic, acousto-optic, free-carrier-dispersion, electro-absorption, and piezo-mechanic, the Pockels or EO effect, present in materials with $\chi^{(2)}$ nonlinearity, is a clear front-runner.

An ideal material platform for electro-optics necessitates a strong interaction between optical and electrical fields, along with minimal optical and electrical losses. It should also be amenable to microfabrication to construct advanced and highly complex systems in scalable fashion. While the quest for the material that meets these stringent criteria continues, the emerging thin-film lithium niobate (TFLN) photonic platform presents a compelling solution. Lithium niobate (LN), having played a pivotal role in optical technology for over seven decades, stands as a cornerstone in various technical applications and scientific disciplines. For instance, its use for optical phase and amplitude modulation has been a workhorse of modern communication technology, thanks to its impressive $\chi^{(2)}$ nonlinearity (~30 pm/V), high refractive index (~2.2), and wide transparency window (0.4-5.5 μm). Its ability to be periodically poled also positions it as a crucial component for frequency conversion in nonlinear and quantum optics. Despite these remarkable properties of LN, its integration into complex chip-scale systems has historically been hindered by the absence of high-quality thin films and the difficulties associated with patterning this material. The recent commercialization of wafer-scale thin-film LN-on-insulator[40–42] and the development of high-quality argon-based physical etching[43,44] have ushered a new era of integrated electro-optics. These advancements have enabled nanoscale optical confinement resulting in enhanced EO interaction, while maintaining low optical losses. Brief yet storied history of the burgeoning field of TFLN photonics, has been detailed in several comprehensive reviews in Ref. [45–47].

The essential component of TFLN electro-optics is the high-performance EO modulator: it allows electronic control of all degrees of freedom of light, including amplitude[1,5], phase[48], frequency[3], and polarization state[49], as well as its temporal[50] and spatial profile[51]. TFLN-based modulators feature a very low half-wave voltage $V_\pi$ (voltage required for a $\pi$ phase shift of light, ~1V), low optical propagation loss ($\alpha \sim 1$ dB/m), and high EO bandwidth (BW$\sim$100 GHz) [1,5,43,52]. These attributes are the consequence of a strong EO interaction, long photon lifetime, and low microwave losses, respectively. In comparison to other photonic platforms, such as silicon and indium phosphide, which may excel in one or two metrics, the simultaneous achievement of these three attributes represents a substantial advantage of TFLN photonics. From a physics perspective, TFLN *enables the strongest coupling between optical and microwave modes, setting it apart from all other EO platforms*. From an application perspective, the combination of low $V_\pi$ and high EO bandwidth results in an energy-efficient and high-speed control. Meanwhile, minimal optical loss allows for the integration of many components and/or functionalities in series within the same circuit, allowing far more operations to be performed on light than any other platform, thus offering unprecedented system performance and capabilities.

Another important component of TFLN photonics are EO resonators, which can trap photons for periods of time (photon lifetime ~ 10 ns for cavities with quality factors Q ~ $10^7$) [43]. This duration significantly surpasses the operation time required to extract or inject the photon into and out of a cavity using the EO effect (~ 0.1 ns, Box 1), thereby allowing many EO operations to be performed within one photon lifetime. Furthermore, even more scientifically interesting and technologically pertinent functionalities emerge when two or more cavities are coupled together [3,6–8,13,18–20]. Such configurations enable the interaction of multiple reconfigurable photonic energy levels through strong EO coupling, a feat nearly unattainable in their bulk counterparts. This advancement heralds a new generation of complex photonic devices and chip-scale systems (Box 1) that extend beyond the capabilities of traditional EO devices, which typically rely on EO modulation within a single waveguide and/or cavity. Moreover, combining the EO effect in TFLN with its inherent optical $\chi^{(2)}$ and/ or $\chi^{(3)}$ nonlinearity, as well as piezoelectricity, offers many additional opportunities. These unique features on TFLN have underscored the need for more advanced mode-coupling formalisms in electro-optics, beyond the conventional formalism based on EO tensor and index change (Box 1).

This review aspires to present a comprehensive introduction to TFLN electro-optics, encompassing principles, formalisms, state-of-the-art technical developments, and applications. We hope this review will complement other reviews[45–47,53–56], which provide a broad overview of TFLN devices[45–47], or are focused on nonlinear optics[53], and traditional EO modulators[54–56]. Thus, we hope our work will provide valuable insights into novel EO devices and systems on TFLN, and foster a deeper understanding in this rapidly advancing field. This is an exciting time for TFLN electro-optics, as it is transitioning from novel device-level demonstrations to advanced system level applications.

## Basic principles of electro-optic modulation

**EO (Pockels) effect**

Electro-optics is the study of the interaction between electrical and optical fields through the EO effect. The most basic physical picture for the EO effect can be visualized as a static electric field inducing a constant change in the refractive index $\Delta n$ of the material (Fig. 2a). The index change $\Delta n$ is conventionally described using the electro-optic tensor R with element $r_{ij}$:

$$\begin{bmatrix} \Delta\left(\frac{1}{n^2}\right)_1 \\ \Delta\left(\frac{1}{n^2}\right)_2 \\ \Delta\left(\frac{1}{n^2}\right)_3 \\ \Delta\left(\frac{1}{n^2}\right)_4 \\ \Delta\left(\frac{1}{n^2}\right)_5 \\ \Delta\left(\frac{1}{n^2}\right)_6 \end{bmatrix} = R \begin{bmatrix} E_x \\ E_y \\ E_z \end{bmatrix} = \begin{bmatrix} r_{11} & r_{12} & r_{13} \\ r_{21} & r_{22} & r_{23} \\ r_{31} & r_{32} & r_{33} \\ r_{41} & r_{42} & r_{43} \\ r_{51} & r_{52} & r_{53} \\ r_{61} & r_{62} & r_{63} \end{bmatrix} \begin{bmatrix} E_x \\ E_y \\ E_z \end{bmatrix},$$

in which the $\Delta\left(\frac{1}{n^2}\right)_k$ with $k = 1, \cdots, 6$ is the index change in the format of index ellipsoid in conventional nonlinear optics (see e.g. Ref [57] for details). Consequently, light propagating through this material undergoes an additional phase shift $\Delta\phi$ proportional to the index change $\Delta n$ and the propagation distance $L$.

**EO modulation**

When a temporally varying electric field, for example a sinusoidal microwave signal, is applied, the induced phase shift $\Delta\phi$ oscillates with the amplitude of the microwave signal, resulting in the generation of new frequency components of light, with frequency intervals matching that of the microwave signal[58,59]. This process, known as *EO modulation*, is a fundamental mechanism in electro-optics. Physically, the EO modulation involves the interaction of the microwave field with the optical field, where frequency of optical photons is changing through the absorption or emission of microwave photons. From an application standpoint, EO modulation serves as the bridge between microwave and optical domains, playing a pivotal role in the global optical communication network. For example, using this process, analog microwave signals can be imprinted on an optical carrier, as it is done in the fields of microwave photonics, or digital information can be encoded in both amplitude and phase of the optical signal, as it is done in optical communications. Conversely, optical signals with differing frequencies can interact via the EO effect, generating microwave signals that can be used to excite or control an electronic system.

A detailed formalism of EO modulation can be found in Box 1. Conventionally, the EO modulation is used to perform rapid modification of degrees of freedom of light such as temporal profile (e.g. by switching light on and off) and spectral content (e.g. by generating sidebands). In Box 1, we also discuss the mode-coupling formalism in which the EO modulation is represented as a coupling between different modes/energy levels and excites EO transitions of photons, which needs to be considered due to the strong EO coupling regime offered by TFLN.

**TFLN photonics devices: cross section and material stack**

In TFLN photonics, the basic device structures are the optical waveguides used to confine the light, and the electrodes to support the microwaves. The film of LN used to support such waveguides and electrodes are created via "smart cut" technology by slicing the bulk LN along a specific crystal axis (Fig. 2b). LN is birefringent material: it has two different refractive indices along different axis. Typically, the x and y axis have an ordinary refractive index $n_0$ and z-axis has extraordinary index $n_e$. The largest EO tensor element is $r_{33}$ (~31 $\mathrm{pm/V}$), which requires applying the electric field along the crystal z-axis. Therefore, the electrode used to support the microwave field is designed to have the electrical field penetrating the slab along the z-axis of TFLN. The most commonly used films of LN can be classified into two types: x-cut and z-cut. The "x-cut" indicates that the thin-film of LN is precisely cut normal to the crystal x-axis, thus extending the film in the y-z plane of the LN crystal. Similar arguments apply to z-cut. In x-cut TFLN, the waveguide is ideally defined along the y-axis of LN, and parallel to it a pair of metal electrodes is introduced to form a capacitor, or for high-speed application a microwave transmission line (see Box 3 for microwave engineering of EO modulation). The signal and ground metal of the transmission line support an electrical field along the z-axis (Fig. 2c). This configuration may not be utilized nor possible in all designs, e.g. complex resonant structures, and in this case, only the field projection onto the z-axis contributes to effective modulation. In z-cut TFLN, the transmission line is designed so that the signal is on-top of the waveguide, leading to an electrical field perpendicular to the film plane (Fig. 2c).

Additionally, another viable option is the hybrid form of TFLN, which involves the creation of a waveguide through heterogenous film-bonding[60–65] (Fig. 2c) or rib loading[66–76] (see Fig. 2c). The hybrid TFLN platforms avoid etching LN, which alleviates concerns of possible Li+ contamination. This has led to extensive research on integrating TFLN with mature silicon and silicon nitride integrated photonics[60–63,66–76]. Both chiplet/coupon-level[60–63] and wafer-scale bonding of TFLN-on-Si or -SiN has been demonstrated[64,65]. Recently, high-quality monocrystalline Si-LNOI wafers have also been Id[68,69]. These hybrid platforms typically have weaker mode confinement in the LN layer and therefore somewhat degraded EO efficiency. Nonetheless, with optimized mode engineering, state-of-the-art hybrid EO devices are exhibiting comparable performance to the monolithic ones that may already satisfy most practical applications[4,60,77–79].

While in this review we focus on monolithic (etched) x-cut TFLN waveguides, we note that our discussions and conclusions qualitatively apply to other system configurations.

## Major classes of EO devices on TFLN

The three major classes of TFLN EO devices are those based on waveguides, cavities (resonators), and coupled cavities (Fig. 2d). The most fundamental structure is the waveguide-based phase modulator (PM)[48], which allows for precise phase control of the light passing through it. By placing one PM in each arm of an on-chip Mach-Zehnder interferometer (MZI), an amplitude modulator (AM) can be realized. This device leverages the interference between two paths that light can take to exert control over the amplitude of light[1,4,5,52]. Single-cavity devices can be achieved by placing a PM inside a cavity[2,14] . In this configuration, the phase change effectively alters the free spectral range (FSR) of the cavity, leading to a change of the frequency of the resonance. Note that the change of FSR due to EO effect is typically much smaller than the FSR itself, thereby in most of cases, only resonance frequency modulation is observed under an external phase modulation. Alternatively, one can embed an AM in a cavity, allowing control over the resonator’s loss. It is worth noting that phase-modulated and amplitude-modulated cavities manipulate the real and imaginary components of the mode’s energy. More advanced devices enabled by TFLN are coupled-cavity devices[3,6,8,13]. They typically consist of two or more evanescently coupled cavities. Each cavity supports discrete modes – energy levels – each with its own loss or gain (e.g. from ion-doping or parametric gain). The advancement of coupled-cavity devices on TFLN lies in the EO modulation, which enables photonic EO transitions between different energy levels. These levels and their associated loss/gain are highly reconfigurable. Consequently, various coupled-cavity devices with new functionalities and high performance have been realized over the past several years. Section 3 discusses specific EO devices in more detail.

## Fundamental rules for efficient electro-optic coupling/transition

While conventionally EO modulators are treated as switch/pulse generator in temporal domain or sideband generator in frequency domain, the EO modulation on TFLN for advanced EO devices/functionalities (e.g. cascaded frequency shifting, topological transport in cavity-based devices, EO frequency comb generations, EO synthetic dimensions, etc) can be more conveniently described as a process of EO coupling/transition between different modes (i.e. energy levels). The design of these systems using the EO mode-coupling formalism, should account for three fundamental rules: energy conservation, momentum conservation, and selection rules.

*Energy Conservation.* This rule is straightforward. For instance, if the input light is at a frequency $\omega_{L1}$ and the goal is to generate a frequency component $\omega_{L2}$, a microwave signal with a frequency of $\omega_m = |\omega_{L2} - \omega_{L1}|$ is required. In the case of a resonant structure, for the modulation process to be efficient, the frequencies $\omega_{L1}$ and $\omega_{L2}$ must also match the frequencies of the two individual resonant modes of the device, $\omega_1$ and $\omega_2$.

*Momentum Conservation.* In all-optical three-wave mixing, the momentum conservation (also called phase matching) between three participating optical modes is an important consideration. In electro-optics, one of the optical modes is replaced with electromagnetic signal of a lower frequency, typically in the microwave range. Given the relatively long wavelength of microwaves (ranging from centimeters to millimeters), their momentum ($k_{MW} = 2\pi/\lambda_{MW}$) is typically small. As a result, the momentum conservation considerations are only relevant at high frequencies (>10 GHz) in which the $k_{MW}$ is not negligible. Additionally, momentum conservation becomes significant only when travelling wave configuration is used for the microwave electrode (see Box 3 for details), as standing wave microwaves lack net momentum. Consequently, this momentum conservation effect is particularly relevant for waveguide-based traveling-wave EO modulators. It is worth noting that, in modulator design, this momentum consideration is often referred to as velocity matching (refer to Box 3 for details).

*Selection Rule.* Similar to atomic physics, where an atomic transition is only allowed when an external drive induces a non-zero off-diagonal coupling between the two states in the Hamiltonian, EO transitions adhere to a similar selection rule: there should be a non-zero overlap between the initial and final optical modes in the presence of an EO interaction. For instance, consider the photonic two-level system that is formed by a double-ring coupled-cavity (see Box 1 for details), of interest for realization of frequency shifters and beam splitters[3], quantum transducers[18,19], etc. Here, EO modulation applied to the device can drive a transition between the two eigenmodes, in this case symmetric (S) and anti-symmetric (AS) modes. Note that the phase accumulation due to the EO effect on each ring needs to be different, in contrast to the same phase advance (or delay) on both rings, which will not create mode overlap between the S and AS modes via EO coupling. The same phase advance (or delay) only creates a global phase shift for the S or AS modes (see Box 1 for details). Another example is the cavity EO comb, which employs an EO-modulated racetrack-style cavity in the roundtrip resolved regime (see Box 1 for the definition of modulation regimes). This setup aims to establish coupling between modes separated by a cavity FSR. These different modes experience different overall phase accumulation after light makes one roundtrip in the cavity. Note that in the roundtrip-resolved regime that we define as $\omega_m$~FSR or $\omega_m \gg FSR$, the microwaves cannot be regarded as unchanged during one roundtrip of light. Consequently, the electrode is designed such that, at a fixed point in time, the upper arm of the cavity has phase advance (delay) while the bottom arm has phase delay (advance). This configuration ensures that light experiences the same phase advance (or delay) in both arms because the sinusoidal microwave voltage reverses its sign after half a roundtrip. Thus, based on the specific functionality of the devices, the selection rule must be deliberately tailored to create mode-overlapping coupling and needs to be carefully considered for different modulation regimes.

## Electro-optics enabled by thin-film lithium niobate

### Phase and amplitude modulation

EO phase shifters/modulators are the simplest EO devices. They can be realized by passing a waveguide through a pair of electrodes[80,81,48] (see Fig. 2a, b, c). Embedding PMs in an interferometer or resonator can implement an AM. The most widely used form of AM is the traveling-wave Mach-Zehnder modulator (MZM) (Fig. 2d, e, f), where the two arms of a Mach-Zehnder interferometer pass through the two gaps of a coplanar waveguide (CPW) electrode[1]. In a MZM, since the two arms experience opposite phase shifts, $V_\pi$ is half of that of a phase modulator with the same length. Key FOMs of an EO modulator include insertion loss, half-wave voltage ($V_\pi$), EO bandwidth, linearity, and extinction ratio (only applicable for AMs). These FOM are interrelated, and tradeoffs exist. For instance, a smaller electrode gap reduces $V_\pi$ but may increase metal-induced optical absorption; a longer modulator has smaller $V_\pi$ but leads to smaller EO bandwidth due to microwave loss and possibly imperfect velocity matching. Overall, the EO performance largely depends on electrode design, in which velocity matching, impedance matching, and microwave loss are important considerations (see Box 3). A key advancement in improving the EO bandwidth is the adoption of capacitively loaded CPW electrodes[52,82–84], where microstructured segments are added in the CPW gaps. Doing so prevents current crowding on the edge and reduces microwave loss. One drawback is that the capacitive loading reduces the microwave phase velocity, which requires a low-index handle wafer (such as quartz[52]) or undercuts in the high-index Si substrate[82] to maintain velocity matching.

State-of-the-art TFLN MZMs have achieved $V_\pi$ of ~1 V, extinction ratio > 20 dB, 3-dB bandwidth > 100 GHz, and total insertion loss of a few dB (including a large fiber-chip coupling loss and a sub-dB on-chip insertion loss)[5]. Combining two MZMs can realize an in-phase/quadrature (IQ) modulator[85] (Fig. 3g, h), an essential component for coherent communications. With a dual-polarization IQ modulator, a record-high single-wavelength data rate of 1.96 Tb/s was demonstrated[5,86]. Some more recent developments include meander designs of long modulators[87,88], investigation of mm-wave operation[89], improved linearity using a ring-assisted Mach–Zehnder interferometer configuration[90] (Fig. 1i, j), and double-/multi-pass (or loop-back) phase modulators for RF $V_\pi$ reduction (Fig. 1b, c) (see specific applications in Section "Waveguide-electro-optic frequency comb")[12,91].

The transparency of LN at visible wavelengths renders TFLN modulators compatible with technologies ranging from bio- and environmental sensing to AR-VR and quantum information science and technology. Thanks to the wavelength dependence of the EO phase shift $\Delta\phi \sim \lambda^{-1}$ as well as the reduced waveguide-electrode gaps at shorter wavelengths[10,11], AMs with $V_\pi \cdot L$s as low as 0.17 V·cm at 450 nm have been reported[11], with highest reported 3-dB bandwidths (~35 GHz) being limited only by the electrode design[10,92]. Quantum applications of TFLN visible modulators have demonstrated, including  scalable processing single photons from a deterministic quantum emitter[93] and photonic engines integrated with spatial light modulators for fast, free-space control of atomic systems[94].

As mentioned, the TFLN platform supports various micro- and nano-structures that allow more complex modulator designs to be implemented. This includes resonant modulators based on micro-rings[3,95–97], photonic crystals[98,99], as well as Bragg grating-based reflector or Fabry-Perot cavities[100–102]. Resonant modulators have small footprint and low $V_\pi$, but high-Q resonances will limit their operating bandwidth. One potential resolution is to use a design where a ring resonator is coupled to the bus waveguide through an MZM, and modulation affects the coupling instead of shifting the resonance[103]. Some other variations of EO modulators include Michelson-interferometer modulators[104–106], EO polarization controller[107], and compact slow-light MZMs[108].

TFLN modulators offers groundbreaking improvements over previous commercial lithium niobate modulators and devices on other material platforms (see details performance comparison in ref. [1,55]). Further improvements on the $V_\pi$, bandwidth, and insertion loss are possible by increasing the microwave-optical mode overlap, improving the optical loss, velocity matching, and microwave loss (see Box 3 for microwave engineering). Reducing the TFLN modulator size is crucial for large-scale electro-optic systems and are possible by folding the microwave electrode or exploring new types of modulators. With direct industrial applications, telecom-wavelength TFLN phase and amplitude modulators are experiencing rapid development, and the ever-expanding literature is difficult to fully capture here. This section only intends to give a brief overview of some representative works and relatively recent progress. For a more thorough and detailed review on this specific topic, readers may refer to Refs. [45,55,56].

## Electro-optic frequency comb generation: waveguide-based

The EO frequency comb is generated by microwave driven modulators, resulting in sideband generation where each sideband represents one comb line. It offers electrical control over the comb dynamics and maintains intrinsic phase coherence among all comb lines. A comprehensive review on EO combs can be found in Ref. [109]. Traditional EO combs are typically waveguide-based. In this setup, light passes through a PM, and the output light can be expressed as $E_{out} = E_0 e^{i\omega_o t - ikz} e^{i\beta \cos(\omega_m t + \phi)}$. The term $e^{i\beta \cos \omega_m t}$ is the generation function of a Bessel function. Therefore, one can directly expand $e^{i\beta \cos \omega_m t}$ to obtain the analytical solution for the output field: $E_{out} = E_0 e^{i\omega_o t - ikz} \sum_n J_n(\beta) e^{in(\phi + \frac{\pi}{2})} e^{in\omega_m t}$. Building upon this, one can cascade several PMs or AMs together to obtain a broader waveguide EO comb[110], in which the use of AMs flattened the comb shape.

The field of waveguide EO combs has been rapidly growing in recent years. Frequency combs with more than 40 spectral lines spanning 10 nm have been demonstrated using only a single PM[48] with $V_\pi = 3.5 - 4.5$ V at $5 - 40$ GHz. A microwave driving power of ~3.1 W is used, corresponding to a modulation index $\beta \sim 4\pi$ [48]. Building upon this breakthrough, flat-top waveguide EO combs on TFLN were first demonstrated by cascading an AM with a PM, as shown by Yu et al[9] and Xu et al[111]. For example, in Yu et al[9], a total of 67 comb lines with a 12.6 nm span were demonstrated by driving the device with 3.8 W microwave power. This work utilized a loop-back structure for the PM to reduce the $V_\pi$ of the PM to $2 - 2.5$ V. Loop-back geometry, however, creates a weak resonance that makes the $V_\pi$ frequency dependent (multiples of ~3 GHz in frequency combs reported by Yu et al[9]). Although bulk system can achieve similar comb line number and spans, it requires a cascade of one AM with three PMs. TFLN waveguide EO combs significantly reduce the system complexity. One drawback of current EO combs is the need for multiple high power microwave sources. Addressing this challenge could be groundbreaking for waveguide EO combs. These combs are ideally suited for applications that require good spectral flatness and do not necessitate wide span, such as data communications.

## Synthetic electro-optic crystal generation

The concept of a photonic synthetic dimension in the frequency domain is rooted in the arrangement of distinct optical frequencies to construct a lattice. Nonlinear processes, such as EO modulation, facilitating lattice coupling. This approach holds promise for simulating complex physical systems and exploring novel physics, due to the high frequency and wide bandwidth of light, along with reconfigurability, scalability, and the capability to control gain and loss. These attributes have enabled the exploration of non-Hermitian[112,113], high-dimensional[14,114], and topological systems[115–119], to name some examples, as well as Moire lattices[120], which can be challenging to study using other methodologies. The simplest configuration of a synthetic EO crystal involves a single EO modulation applied to a cavity with microwave frequency equal to the FSR (Fig. 2l). This is formalized by considering the EO cavity in the roundtrip resolved regime (see Box 1). Advanced structures that involve multiple modulation tones or coupled cavities can be extended in this formalism (see Box 1). Although simple in concept and geometry, this field has shown great potential and excellent experimental progress in fiber-optics systems[113], but has been limited by scalability and high insertion loss challenges inherent to this approach. TFLN holds promise as a catalyst for advancing this field: the strong EO enables strong coupling rates and low loss between lattice points, orders of magnitude beyond the loss rate of each lattice point (EO coupling $\Omega \sim 10$ GHz, optical loss rate $\kappa \sim 100$ MHz), a regime that is inaccessible for other photonic platforms. These features suggest large-scale and intricate synthetic crystals can be realized.

Recently, high-dimensional frequency crystals, extending up to four dimensions, were demonstrated using a single EO cavity on TFLN, in addition to measurement of density of states and a coherent random walk[14] (Fig. 2m). Instead of folding a 1D lattice to achieve higher dimensionality, as is done conventionally[119,121,122], rather the high-dimensionality is achieved by using several different microwave drives with their frequencies closely matching (mutually detuned by 1 MHz) the cavity FSR (~10 GHz). As a result, each microwave drive provided an individual dimension for photons to hop, leading to each dimension encompassing over 100 lattice points. Furthermore, a crystal with a frequency domain mirror, i.e. a device capable of reflecting optical energy propagation along the frequency dimension, was proposed and experimentally demonstrated[13]. This was accomplished using both coupled-cavity and polarization crossing. The mirror exhibited a reflectivity exceeding 0.9999, showcasing interference of forward and backward waves, along with the observation of trapped states in the frequency domain. More recently, exploration of synthetic dimensions using TFLN has gone quantum and used to explore correlations between entangled photons and multilevel Rabi oscillations, in which quantum frequency conversion was employed to create a second dimension[15].

**Electro-optic frequency comb generation: cavity-based**

The simplest configuration of a cavity EO comb involves placing a phase modulator (PM) inside a high-Q cavity. Light circulates within the cavity, passing through the PM multiple times, which, as discussed in Sec. 2 (selection rule), enables broadband EO comb generation. Tight confinement of the waveguide in TFLN, along with a high-performance PM, facilitates this configuration while maintaining a high-Q cavity. This is a significant advantage compared to bulk systems where the insertion loss of bulk components is typically large. The underlying mechanism of single-cavity EO combs can be modeled using roundtrip-resolved modulation (see Box 1 for details), the same model used to study EO synthetic crystals. In 2019, a single-cavity EO comb was successfully demonstrated on TFLN (Fig. 2h), achieving a span one order of magnitude larger than previous EO combs[2,123]. However, despite this impressive span improvement, the conversion efficiency of this type of comb remains limited, approximately 0.3% [6]. The low conversion efficiency arises from the cavity becoming strongly under-coupled when driven with a strong microwave tone. The microwave drive introduces a broadband comb generation process, acting as an effective intrinsic loss $\kappa_{MW}$ for the pump mode[6]. This behavior is analogous to the generation process for other types of frequency combs, such as Kerr combs. On TFLN, the effective loss rate due to the microwave drive ($\kappa_{MW}$) is about 100 times higher than the external ($\kappa_e$) and intrinsic ($\kappa_i$) loss rates of the cavity (i.e. $\kappa_{MW} \sim 10$ GHz, $\kappa_e \sim \kappa_i \sim 100$ MHz for $\sim 1$ million quality factor [6]). This means that $\kappa_e \gg \kappa_i + \kappa_{MW}$. Simply increasing the external coupling to achieve critical coupling with the microwave drive results in high loss for resonance modes, reducing the overall bandwidth.

This issue has been addressed by implementing the concept of GCC proposed in Ref. [3], and discussed more in the following section, to a coupled-resonator EO comb (Fig. 2g, i). This led to a 100-fold improvement in efficiency (30%) and a 2.2-fold improvement in span (132 nm)[6]. Another challenge with cavity EO combs is achieving a flat-top comb shape. Recent work has shown that a flat-top cavity EO comb can be achieved by using a boundary as a frequency domain mirror[13] (see previous section). With the significant progress in TFLN-based cavity EO combs, it is anticipated that they will become valuable broadband and stable comb sources, enhancing applications in telecommunications, metrology, astronomy, spectroscopy, and more. It should be pointed out that, cavity-based EO comb has the limitation on the microwave power consumption as well as the tunability of the repetition rate due to the resonant enhancement condition. TFLN has provided a versatile platform for cavity-based EO comb with novel structures, unlocking possibilities to further improving the bandwidth, conversion efficiency, and spectral flatness. Finally, it is important to note that, cavity-based EO comb also generates femtosecond pulse source[6] as conventional EO comb, benefiting applications on ultrafast optics.

**Frequency conversion: general critical coupling and photonic molecules**

Microwave-enabled control of light holds a great promise for applications requiring ultrafast operations, such as frequency conversion, sensing, and switching, due to the compatibility with ultrafast electronics and the stability of microwave signals. However, this typically necessitates a photon lifetime significantly longer than the time required to transition between states, which was difficult to achieve in other photonic platforms. TFLN overcomes this limitation by simultaneously featuring strong EO coupling on the order of gigahertz, and high-Q cavities corresponding to low loss rates of 10-100 megahertz (as defined in Box 1). Consequently, TFLN now facilitates EO transitions between modes of different energy in both single and coupled cavities. For instance, within a single cavity, applying a microwave with a frequency equal to the FSR creates EO transitions across a synthetic frequency lattice. In a coupled-cavity system, microwaves can induce transitions between several hybrid modes, if energy conservation, phase matching, and selection rules are satisfied (as explained in the section on "rules for efficient electro-optic transition").

Recently, a photonic two-level system (TLS) with EO transition was demonstrated using two coupled-cavities on TFLN[3,8] (Fig. 2a). This system can induce EO coupling (with $\Omega$~5 gigahertz) between the S and AS modes of the coupled-cavity (see Box 1 and the selection rule in section 2), leading to Rabi oscillations between the two energy levels. As an optical TLS, this device is well-suited for GHz-scale frequency control, a critical energy band for communications and computing.

Typically, this EO coupling results in energy oscillations between the two frequencies, which is unsuitable for efficient energy conversion. To address this, the concept of general critical coupling (GCC) has been proposed for efficient energy conversion[3] (see Box 2 for detailed introduction). Combining GCC with the TLS coupled to a continuum, e.g. a bus waveguide, enables novel EO frequency shifters and beam splitters (Fig. 2b). Record-breaking frequency shifting performance was demonstrated, with over 99% shift efficiency (shift ratio), 0.45 dB on-chip insertion loss, and a shift frequency range of 10-30 gigahertz, all achieved using only a continuous single-tone microwave[3]. Furthermore, a concept of cascaded frequency shifting is introduced, and a cascaded frequency shifter (a triple-coupled-cavity device) is demonstrated, achieving a four-step cascade to a 120-gigahertz frequency shift using a 30-gigahertz microwave[3]. Here, energy flows through a ladder of energy levels without any back reflection. This device showcases the potential to access energy bands exceeding 100 gigahertz using only a low-frequency microwave source. High-performance frequency shifting and beam splitting operations at GHz band have the potential to impact applications in communications[21,124], classical information processing[35,39,125,126] as well as quantum computing[127–132]. Finally, another type of triple-coupled-cavity system has been demonstrated, forming a three-level system[7] (Fig. 2c), and was used to realize an optical isolator. Nearly 40 dB of optical isolation is achieved using approximately 75 milliwatts of microwave power These novel EO transitions in coupled-cavity systems offer a fully reconfigurable photonic system with ultrafast control, revitalizing the applications of EO systems. Further increasing the mode overlap and the cavity quality factor can significantly improve the coupling regime and extending to large-scale EO cavity array can enable advanced complex system with collective dynamics.

**Frequency conversion: spectral shearing**

Electro-optic spectral shearing is another approach for altering the frequency of light. It enables a substantial shift in frequency, typically on the order of ~100 GHz, and requires only a single-tone microwave driving signal. The underlying principle of spectral shearing lies in the fact that a linearly ramped phase modulation $\Delta\phi = kt$ is equivalent to a frequency shift of light: $E_{out} = E_{in}e^{ikt} = E_0 e^{i\omega_o t - ikz}e^{ikt} = E_0 e^{-ikz}e^{i(\omega_o + k)t}$. It's important to note that while spectral shearing can achieve a very large frequency shift beyond the microwave driving frequency (in the case of sinusoidal microwave drive), it only works for pulsed light. The duration of pulse needs to be shorter than the linearly rising (or falling) part of the sinewave. When using a sinusoidal modulation, the phase modulation applied $\Delta\phi = \beta\sin(\omega_m t + \Delta\varphi) \approx \beta\omega_m t\cos\Delta\varphi$, in which $\Delta\varphi$ is the relative phase between the microwave and optical pulse, $\omega_m$ is the microwave frequency, and $\beta$ is the modulation index (see Box 1 for definition) [12]. Here, the frequency shift $\Delta f = k = \beta\omega_m\cos\Delta\varphi$ of the linear ramping phase represents the magnitude of the frequency shift and is determined by the peak-to-peak voltage and the frequency of the used sinusoidal signal[12]. An important figure of merit is the effective shift magnitude ($\Delta f/\delta f$) that is defined as a ratio between the frequency shift that can be achieved $\Delta f$ and the bandwidth of optical pulse $\delta f$ . Consequently, increasing the frequency of the modulation signal to obtain a larger $\Delta f$ (since $\Delta f \propto \beta\omega_m$) does not necessarily help the $\Delta f/\delta f$. Higher frequency modulation leads to shorter periods, necessitating shorter pulse durations (i.e., higher pulse bandwidth $\delta f$). As a result, the ultimate limit on the $\Delta f/\delta f$ determined by the amplitude of the applied phase modulation, which is constrained by the $V_\pi$ of the phase modulator. Note that serrodyne (i.e. sawtooth) modulation[133,134], in which the voltage is repeatedly linearly ramped, mitigates the bandwidth constraint but the required high-bandwidth microwave signal is difficult to achieve, even with high-bandwidth modulators.

Spectral shearing has been successfully demonstrated in bulk systems[135,136] and employed in quantum applications but typically has limited $\Delta f/\delta f < 1$ with $\Delta f \sim 200$ GHz, due to the high $V_\pi$ of bulk modulators. On TFLN, spectral shearing of single photons with a shift frequency of $\Delta f = \pm 641$ GHz ($\pm 5.2$ nm) has been achieved[12], resulting in an $\Delta f/\delta f$ greater than 1. Furthermore, an impressive 8-fold suppression of pulse bandwidth (from 6.55 nm to 0.35 nm) has been demonstrated by applying quadrature phase modulation using the same device. With sub-volt TFLN at visible wavelengths, a spectral shearing with a $\Delta f/\delta f > 7$ has been achieved with $\Delta f = 6$ GHz [10].

**Cryogenic electro-optics**

The EO effect in TFLN persist at cryogenic temperatures, as well. This offers an effective solution for realizing photonic links for superconducting microwave circuits[137] or for on-chip optical routing and processing for optically active ions or defects in the solid-state. Superconducting devices, such as qubits, logic circuits, and detectors, rely on cryogenic operation. As systems scale up, the RF cables required to address these devices will induce unmanageable heat load to the host cryostats. In addition, for constructing a quantum network that connects distant superconducting qubits, direct transmission of microwave photons through RF cables at room temperature is infeasible due to microwave loss and thermal noise. A promising solution is to replace metallic cables with optical fibers, which have low thermal conductivity and are wide bandwidth (THz) and ultra-low loss (< 0.2 dB/km) optical channels. Central to this solution is an efficient and cryogenically compatible mechanism to transduce between microwave and optical signals. Combining a TFLN EO modulator with superconducting electrodes provides a viable solution.

One approach to achieve an ultra-efficient cryogenic EO interface is to use cavity systems[138–140]. This scheme, referred to as cavity EO, is suitable for coherent, bidirectional microwave-optical quantum transduction[141,142]. On TFLN, it can be implemented by replacing the electrode of a double-ring coupled-cavity modulator with a superconducting resonator (Fig. 2d, e, f), with a resonance frequency matching the doublet splitting of the S and AS optical modes (see Box 1 for details)[18–20]. A strong laser pump at the red mode can upconvert a microwave photon into an optical photon at the blue mode. In cavity EO system, the key FOM is the cooperativity, $C = 4g_0^2 n_p/\kappa_m \kappa_o$, which depends on the EO interaction strength $g_0$, microwave loss $\kappa_m$, optical loss $\kappa_o$, as well as the number of pump photons $n_p$ that can be put into the system without inducing quasiparticles in the superconductor, creating noise photons, or overheating the cryostat[142,143]. State-of-the-art cavity EO transduction on TFLN has achieved an efficiency of 1.02% (internal efficiency of 15.2%)[20]. Further improvement requires microwave and optical cavities with higher quality factors, as well as better optical coupling and cryogenic packaging. It is worth mentioning that even with low efficiency, cavity EO transducers can enable heralded entanglement between distant superconducting systems[144]. Pumping the cavity EO device at the blue hybrid mode can generate pairs of microwave and optical photons via spontaneous parametric down conversion. The optical photons can be sent to a distant location to establish entanglement with another pair source through the DLCZ protocol. Similarly, difference frequency generation between two optical fields can be used to generate microwave signals inside the cryogenic environment, that can be used to control superconducting qubits[145].

Cavity systems enhance EO interaction at the expense of operating bandwidth. For applications such as rapid single flux quantum (RSFQ) circuits, superconducting nanowire single-photon detectors, and frequency-multiplexed superconducting qubits, a broadband, low $V_\pi$ modulator is desired. This can be achieved using a long traveling-wave modulator with superconducting electrode, which has negligible ohmic loss and breaks the tradeoff between length and EO bandwidth. Recently, a superconducting (niobium) TFLN MZM with effective modulation length of 1 m and a $V_\pi$ as low as 42 mV was demonstrated. Such device (a 20-cm-effective-length device with 230 mV $V_\pi$ and 17 GHz bandwidth) is used to transmit RSFQ signals (5 mV $V_{pp}$) at 4 K to room temperature through telecom optical fibers[146].

## Emerging applications

With the recent rapid advancements in EO devices on TFLN, the field is transitioning from focusing on individual devices to comprehensive system-on-a-chip approaches. While there are many expected applications of this platform, including those in telecom, datacom, and microwave photonics, here we focus on several emerging applications and future opportunities.

### Photonic computing and accelerators

Tightly integrated and power-efficient electronic and photonic systems can enable important advances and push the state of the art in optical communication systems. Furthermore, we envision such an integrated electronic-photonic system using TFLN will enable breakthroughs in optical computing as well as microwave photonics. With rapid advancements in AI and development of deep neural networks, CPUs and GPUs alike are challenged by impending bottlenecks in memory and speed while maintaining small form factor and low power consumption. Recently, the photonics community responded to these challenges with novel architectures, specialized towards accelerating the training and inference stages of such networks. The pioneering work of Shen et al. demonstrated audio recognition using a silicon-photonic chip[147]. Following this demonstration, many photonic neural network accelerators emerged, including processors based on frequency combs[148], phase-change materials[149], current-controllable optical attenuators[150], Mach-Zehnder interferometer meshes[151], vertical-cavity surface-emitting laser arrays[152], optical diffraction[153], and in combination with communications[154], further pushing the boundaries of computational accuracy, speed, and power consumption. Recognizing the impact, the TFLN platform promises to provide orders-of-magnitude advantages in aspects of speed and power consumption, and might play a key role in finally bringing photonic neural networks to market. The heterogeneous integration of lasers[155–165] on TFLN provides high-power data channels, while EO[2,6,123] and Kerr[166–168] frequency combs provide potential for massively parallel frequency-division multiplexing across hundreds of channels. In each channel, high-speed EO modulators operating at CMOS-compatible voltages[1,4,5] provide unprecedented levels of computational power and reconfigurability due to the analog nature of optical processing, in addition to efficiently encoding and manipulating data with ultra-low power consumption. Finally, the native $\chi^{(2)}$ nonlinearity coupled with high-speed integrated III-V photodiodes[164,169,170] presents opportunity for on-chip nonlinear activation and opto-electronic conversion and detection, closing the loop for all-optical computation. As TFLN continues to mature in the coming years, we might just witness a fully monolithic photonic neural network competitive or even outclassing conventional electronic approaches.

**Quantum photonics**

The TFLN platform will play an important role in quantum photonics. The high carrier frequency of light allows both high-bandwidth quantum communication and, since this energy is far above the thermal background, the measurement of quantum states of individual photons[171–174]. The properties of TFLN, including low-loss, possibility of large scale integration, possibility for photon-photon interactions through EO and/or optical $\chi^{(2)}$, render it an attractive platform for generating and manipulating quantum states of light[175]. Although a single photon nonlinearity is not currently efficient in TFLN[176], the possibility of beam splitting and high-bandwidth EO phase shifting renders it attractive for linear-optical quantum computing[132]. Moreover, combs and modulators allow for such computing protocols in the frequency domain[50,127–131,177–183], taking advantage of the multiplexing ability of light. Indeed, the coupled ring system[3] offers frequency-domain beam splitting without expanding the Hilbert space to using a single phase modulator[184]. High-bandwidth switching offered by TFLN is also a key ingredient for rendering inefficient spatially multiplexed heralded states to be quasi-deterministic, which can be applied to photon sources and quantum communications[185–187].

The all-optical $\chi^{(2)}$ interactions on TFLN, often performed using periodic poling of TFLN[188], further enhances the EO platform. For example, it can be used for photon pair generation, squeezed light generation, and deterministic wavelength conversion of individual photons. These are of interest for photonic quantum computing, realization of quantum interconnects between devices that operate at different energy scales, up-conversion detection[189,190], quantum sensing and computing using highly squeezed states[191–193], and realization of efficient frequency combs via cascaded $\chi^{(2)}$ interactions[194]. Beyond EO and $\chi^{(2)}$, the piezoelectric coupling of TFLN has been harnessed for quantum interconnects between microwave photons and acoustic phonons, with applications in, for instance, transduction or acoustic modulators[195]. Furthermore, TFLN offers embedded rare earth ions[196], which promise quantum emitters, storage, signal manipulation[197], and sensors in addition to the depth of functionalities offered by TFLN itself. Overall, the depth of impact that TFLN has on quantum photonics with electro-optics is tremendous, and only recently we have begun exploring the potential of its impact.

**Active topological photonics and non-Hermitian physics**

Microwave-coupled optical lattices in frequency domain realized on TFLN[3,6,7,13,18,19] are emerging as a promising platform for studying topological photonics, non-Hermitian physics, and microcavity sensing, for example. Combining TFLN with microcavity sensing[198,199] is particularly promising for sensors that are both high-speed and highly sensitive. With the inherent strong piezoelectric interaction on TFLN, EO modulation could add a new control of degree of freedom to conventional acoustic sensors[198,199]. Moreover, EO modulation on TFLN may enable the practical implementations of modulation-induced gauge field for photons[200]. Collectively, these advancements underscores TFLN's unique value in offering additional degrees of freedom for ultrafast tuning for various optical systems, making it an ideal platform for studying optical physics using waveguide and cavities and forging new directions for the manipulation and study of light.

**Nonlinear photonics**

The excellent $\chi^{(2)}$ and $\chi^{(3)}$ nonlinearities of TFLN allow parametric frequency-conversion processes to be combined with electro-optics, opening multiple opportunities for exploring fundamental physics, developing novel devices, and addressing practical applications. From physics perspective, the combined $\chi^{(3)}$-EO [6] and $\chi^{(2)}$-EO [15,201] introduces extra degrees of freedom and interactions for synthetic EO lattice. For example, the coexistence of EO and Kerr effect[6,202] indicates the feasibility of introducing the on-site nonlinear interactions via Kerr effect in the synthetic EO lattice. This allows for the simulation of strong-correlated condensed matter models beyond the single electron approximation. From an applications standing point, integrated nonlinear frequency combs are particularly noteworthy, covering applications such as tunable optical parametric oscillators[201,203,204], ultra-stable microwave[205,206] and millimetre-wave generation[207], optical frequency referencing[208,209] and synthesis[210], astrocomb[211,212], and ranging[26,213]. The potential of integrated nonlinear frequency combs is set to expand by leveraging the multiple strong nonlinearities on TFLN. For example, at device level, TFLN facilitates novel comb generation that may harness the best properties of EO and Kerr effect, such as EO-Kerr comb in the same resonator[6], EO-tunable Kerr combs[214], and cascading EO modulators with Kerr resonators[215]. At system level, the combination of comb sources, modulators, and periodically poled lithium niobate waveguides can benefit these applications. Furthermore, the integration of new gain technologies such as the reflective semiconductor optical amplifier[157] and ion-embedding methods such as Erbium-doping of TFLN[216–218] and Thulium-doped cladding oxides[219], with TFLN is promising. It may lead to exciting avenues in integrated lasers[155–165] at exotic wavelengths and ultrafast, high peak power pulse generators[220] and amplifiers[219,221]. Such advancements may promise the deployment of chips for atomic physics at visible and near-UV[222], fully integrated supercontinuum generation[223], and biomedical imaging[224].

In this review, we have discussed the basic principles and formalisms of electro-optics on TFLN and introduced key research topics in this field. Electro-optics on TFLN has been developed tremendously over the past five years. It is unfolding as one of the most promising solutions for the next-generation of EO interfaces, such as the field of optical communications, microwave photonics, and photonic computing, where efficient and high-bandwidth EO conversion are paramount. It has the potential to continue providing revolutionary EO devices and circuits as well as hybrid circuits that interface with all-optical nonlinearities and acousto-optic interactions. As it matures, we expect TFLN to become one of the most important platforms of integrated photonics.

**Acknowledgement**

This work is supported by: Army Research Office/Dept of the Army (MIT) W911NF1810432; Air Force Office of Scientific Research FA9550-19-1-0376; Office of Naval Research N00014-20-1-2425; Air Force Office of Scientific Research (MIT); FA9550-20-1-01015; NIH/NEI (MGH) 5R21EY031895-02; Defense Advanced Research Projects Agency HR001120C0137; National Science Foundation EEC-1941583; Air Force Research Laboratory (Rigetti) FA864921P0781; Office of Naval Research (Vector Atomic) N00014-22-C-1041; National Science Foundation OMA-2137723; NSF (U Oregon) 2138068; NASA (U Oregon) 80NSSC22K0262; Department of the Navy (MAGiQ); NASA (ICARUS) 80NSSC23PB442; Air Force Office of Scientific Research (MURI UCB) FA9550-23-1-0333; National Research Foundation of Korea; Amazon Web Services Award number A50791; Harvard , FY 22 Spring Dean's Competitive Fund for Promising Scholarship; A*STAR (C230917005) and NRF (NRF2022-QEP2-01-P07, NRF-NRFF15-2023-0005); National Science Foundation Graduate Research Fellowship under Grant No. DGE-1745303

**Box 1 Formalism of electro-optic modulation**

The emergence of TFLN has expanded the landscape of EO devices, including single-cavity- and coupled-cavity-based configurations in addition to the conventional waveguide-based devices. These devices support many resonant modes with distinct frequency/ energy, that can be coupled via EO effect. This process of EO coupling can be understood as removing and/or adding photons from/ to a particular energy level (mode). Typically, the coupling strength Ω induced by EO effect is on the order of ~1-10 GHz[3,6], corresponding to an operation time of ~0.1 ns. This is much shorter than the cavity life time that can be routinely achieved in TFLN, that is on the order of ~1-10ns for resonant modes with quality factor from $10^6$ to $10^7$. This unique combination of large EO coupling strength (i.e. short coupling time) and ultra-low loss (i.e. large photon lifetime) supported by TFLN offers a fertile playground to explore numerous physical phenomena. This proliferation necessitates the development of new formalisms to accurately describe and analyze these increasingly intricate systems. In this section, we will first introduce the conventional formalism used to study EO effect, and subsequently discuss a recently developed approach based on mode-coupling.

**Conventional approach**

The traditional formalism to treat the EO process in LN has been used extensively in the field of nonlinear optics[45,109]. It is based on the concept that an applied voltage introduces a phase shift to the optical field $\Delta\phi = \beta = \pi\frac{V}{V_\pi}$, where $V_\pi$ represents the voltage required for a $\pi$ phase shift. The term $\beta$ is referred to as the modulation index or modulation depth. It is used to quantify the magnitude of the phase shift, essentially indicating how many $\pi$ phase shifts are induced by the applied voltage. Consequently, light transmitted through a waveguide phase modulator, subject to a modulation signal $V_{mod} = V\cos(\omega_m t + \varphi)$, acquires a phase term: $E_{out} = E_{in}e^{i\beta\cos(\omega_m t+\varphi)}$, which can be expanded by Bessel function: $E_{out} = E_{in}\sum_n J_n(\beta)e^{in(\phi+\frac{\pi}{2})}e^{in\omega_m t}$. While conceptually simple, this formalism has complicated description when dealing with systems involves multiple spatial and frequency modes coupled together (e.g. multi-coupled waveguides or cavities with EO modulations), structures/regimes recently achievable on TFLN.

**Mode-coupling approach**

In contrast to using index change to represent the phase shift, the mode-coupling approach considers modulation as a coupling mechanism between frequency modes and involves an analysis of the system's Hamiltonian. Here we discuss the mode-coupling approach for three basic photonic structures: *single cavity, coupled cavity, and waveguide.*

**1. Modulation in a single cavity.**

When it comes to modulation within a single cavity, it is essential to recognize three distinct regimes that require separate considerations. They are defined as follows:

- **Sideband unresolved regime:** $\omega_m < \kappa$, when modulation frequency is smaller than linewidth of the mode.
- **Sideband resolved regime:** $\omega_m > \kappa$ but $\omega_m \ll$ FSR, that is modulation frequency is larger than linewidth but much smaller than the FSR of the cavity.
- **Roundtrip resolved regime:** $\omega_m \sim$FSR or $\omega_m >$ FSR, modulation frequency is comparable with or larger than the FSR of the cavity.

In the case of TFLN, the ultrahigh quality factor resonances typically have MHz-level linewidths that is much smaller than typical modulation frequency that can be applied typically in GHz range. Consequently, we will focus our discussion on the sideband resolved regime and roundtrip resolved regime, only.

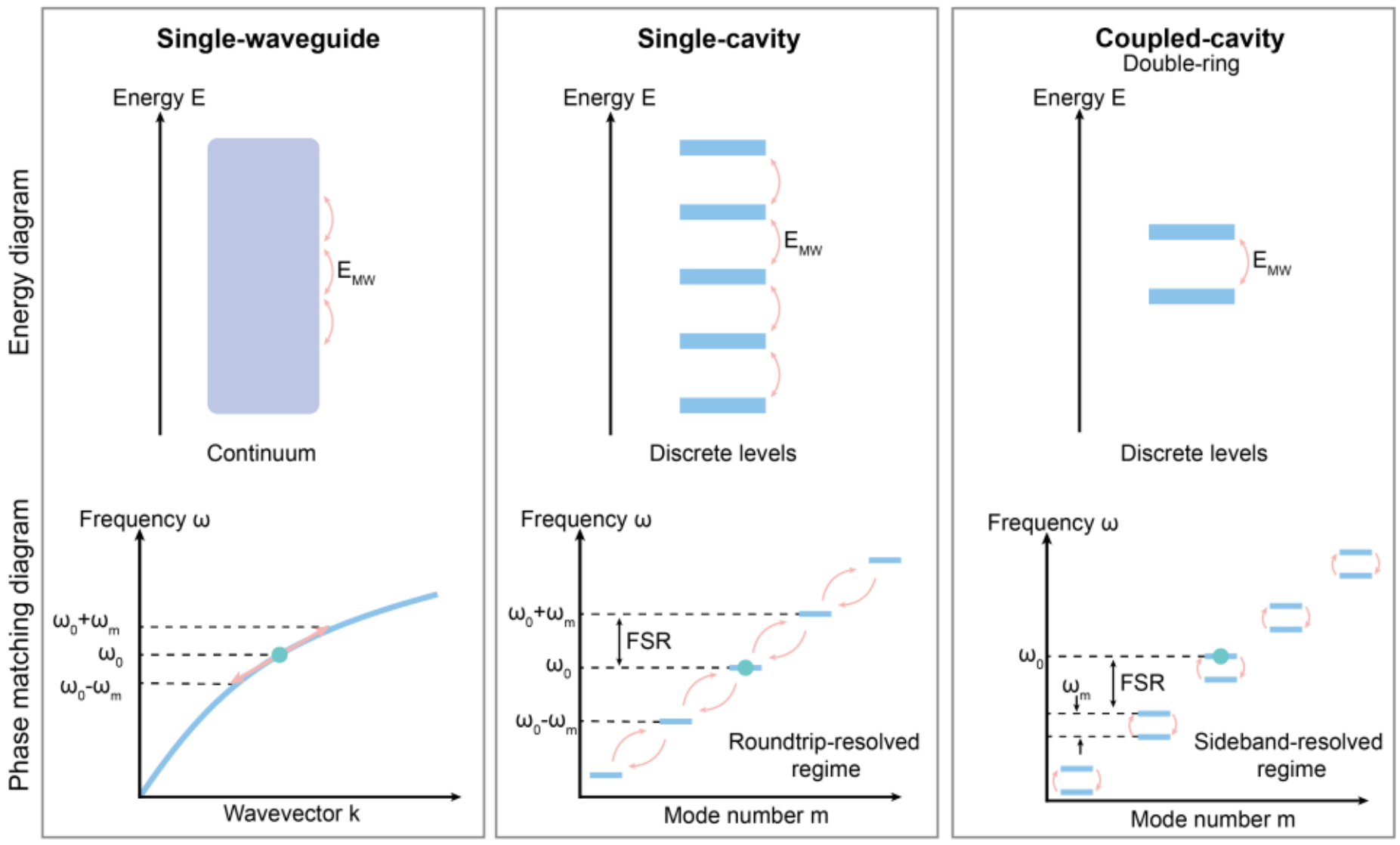

**Box Figure 1** | Energy diagram and phase matching diagram for single-waveguide, single-cavity, and coupled-cavity systems with microwave modulation under different modulation regimes. $\omega_m$, frequency of microwave; $E_{MW}$, energy of microwave photon; FSR, free spectral range. Green dot represents an actual optical photon.

In the **sideband resolved regime**, a cavity incorporating a phase modulator undergoes frequency modulation, expressed as $\omega_0 + \Omega \cos \omega_m t$. This configuration yields the Hamiltonian:

$$H = \omega_0 a_0^\dagger a_0 + (\Omega \cos \omega_m t) a_0^\dagger a_0$$

In which $\Omega = G_V \times V$ represents the EO coupling strength introduced by the microwave, $a_0$ is the cavity eigenmode with frequency $\omega_0$, and $V$ is the peak voltage of the microwave drive. The factor $G_V$ (approximately 0.5 GHz/V in TFLN) is related to the $V_\pi$ of the modulator. It characterizes the frequency shift of the resonance when a 1V DC voltage is applied. Utilizing the Heisenberg-Langevin equation, the equation of motion can be derived as: $\dot{a}_0 = \left(-i\omega_0 - \frac{\kappa}{2}\right) a_0 - i(\Omega \cos \omega_m t) a_0 - \sqrt{\kappa_e} a_{in} e^{-i\omega_L t}$, where $a_{in} e^{-i\omega_L t}$ represents the pump field, $a_0$ denotes the cavity pump mode, $\omega_0$ is the cavity resonance frequency, $\kappa = \kappa_e + \kappa_i$ corresponds to the cavity linewidth, with $\kappa_e$ being the external coupling between cavity and bus waveguide, and $\kappa_i$ signifying the intrinsic loss rate of the cavity. The output field from the bus waveguide is described by $a_{out} = a_{in} e^{-i\omega_L t} + \sqrt{\kappa_e} a_0$.

In the **roundtrip resolved regime**, where the modulation period ($1/\omega_m$) is comparable to the roundtrip time (1/FSR), the cavity's behavior cannot be simplistically viewed as frequency modulation. This is because the concept of cavity frequency is established through the roundtrip interference effect. In this scenario, the EO coupling directly affects a set of optical modes $\{a_n\}$ ($n = -N, \cdots, N$ with 2N+1 being the total number of modes considered) separated by the FSR. The Hamiltonian is given by:

$$H = \sum_{n=-N}^{N} \omega_n a_n^\dagger a_n + (\Omega \cos \omega_m t)(a_n a_{n+1}^\dagger + a_n^\dagger a_{n+1})$$

One can derive the equation of motion for $\{ a_n \}$: $\dot{a}_n = \left(-i\omega_n - \frac{\kappa}{2}\right) a_n - i\Omega \cos \omega_m t \, (a_{n+1} + a_{n-1}) - \sqrt{\kappa_e} a_{in} e^{-i\omega_L t} \delta_{n,0}$. Details of solving such systems can be found in ref. [6].

It is worth noting that, in conventional approach, the coupling coefficient $k$ between waveguide and cavity, roundtrip loss $\alpha$, modulation index $\beta$ are all unitless, while the $\kappa_e$, $\kappa_i$, and $\Omega$ is in the unit of hertz in the Hamiltonian. The relationship between these is $k = 2\pi \frac{\kappa_e}{\text{FSR}}$, $\alpha = 2\pi \frac{\kappa_i}{\text{FSR}}$, $\beta = 2\pi \frac{\Omega}{\text{FSR}}$ (e.g. $\kappa_e = 2\pi \times 100$ MHz, FSR $= 2\pi \times 10$ GHz $\rightarrow k = 0.06$)[6].

## 2. Modulation in a coupled cavity.

For simplicity here we discuss the case of two coupled cavities with applied modulation in sideband resolved regime (Fig. 3a). More intricate configurations of coupled-cavity structures typically build upon or combine elements from this foundational case or incorporate the modulation in the roundtrip resolved regime. The Hamiltonian describing two coupled cavities in the sideband resolved regime is given by: $H = (\omega_0 + \Omega\cos\omega_m t)a_1^\dagger a_1 + (\omega_0 - \Omega\cos\omega_m t)a_2^\dagger a_2 + \mu(a_1^\dagger a_2 + a_1 a_2^\dagger)$, in which $\mu$ represents the strength of the evanescent coupling between two rings. This system can be simplified by applying a basis transformation: $c_1 = (a_1 + a_2)/\sqrt{2}$ and $c_2 = (a_1 - a_2)/\sqrt{2}$. In contrast, this yields a Hamiltonian akin to a photonic two-level system, with an interaction term analogous to Rabi oscillation:

$$H = \omega_1 c_1^\dagger c_1 + \omega_2 c_2^\dagger c_2 + \Omega\cos\omega_m t\,(c_1^\dagger c_2 + c_1 c_2^\dagger)$$

where $\omega_1 = \omega_0 - \mu$ and $\omega_2 = \omega_0 + \mu$. If both phase advance (or delay) is applied on $a_1$ and $a_2$, it only creates a global phase at $c_1$ or $c_2$. One can see that two ring with different phase direction changes $(a_1 + a_2)$ towards $(a_1 - a_2)$ and vice versa, thus satisfy the selection rule. Details of solving such a system can be found in the Methods of ref. [3]. It is worth noting that only one ring is coupled to the waveguide, therefore the output field is $a_{out} = a_{in} + \sqrt{\kappa_e}a_1 = a_{in} + \sqrt{\kappa_e}\frac{1}{\sqrt{2}}(c_1 + c_2)$.

**3. Modulation in waveguide.**

In the waveguide, the spatial dimension z along the propagation direction assumes the role of time in the equation governing the EO-modulated cavity. For light $a_0 e^{i\omega_L t - ikz}$ with frequency $\omega_L$ passing through a single waveguide phase modulator, it gradually couples to other frequency modes $a_n$ with frequencies $\omega_o + n\omega_m$ $(n = \cdots, -1, 0, 1, \cdots)$ as it propagates along the phase modulator:

$$\frac{\mathrm{d}}{\mathrm{d}z} a_n e^{ik_n z - i\omega_n t} = -i\Omega\cos(\omega_m t - k_m z)(a_{n-1}e^{ik_{n-1}z - i\omega_{n-1}t} + a_{n+1}e^{ik_{n+1}z - i\omega_{n+1}t})$$

Similar techniques involving the rotating frame can be employed to solve these equations. It is important to note that this represents a traveling optical wave in a waveguide, where phase matching considerations come into play as indicated by this equation.

## Box 2 General critical coupling for efficient energy transfer

In optical systems, resonant modes with critical coupling conditions are commonly employed to enhance optical interactions (i.e., waveguide-cavity external coupling $\kappa_e$ = intrinsic loss $\kappa_i$). Nevertheless, in intricate optical systems, there may be multiple loss channels beyond the external and intrinsic losses. In particular, some optical mechanisms could serve as a loss channel for specific modes (e.g., the comb generation process acting as a loss channel for the pump mode, as energy is extracted from the pump and converted into the comb). Consequently, to facilitate efficient energy transfer and manipulate energy flow, it becomes essential to incorporate the concept of general critical coupling (GCC). Within this framework, in addition to the conventional losses stemming from coupling to the continuum ($\kappa_e$ due to coupling to waveguide, $\kappa_i$ due to scattering to environment), coupling to the resonant mode also introduces an effective loss.

This process can be modeled as a two-level system problem, in which one resonant mode $a$ has conventional $\kappa_e$ and $\kappa_i$ and a second mode $b$ with total loss rate $\gamma$ is coupled to $a$ through an interaction with coupling strength $\mu$ (this $\mu$ can be evanescent coupling between two rings, EO transition from microwave drive, or nonlinear interaction such as second harmonic generation). This process can be modeled as

$$\dot{a} = \left(-i\Delta_a - \frac{\kappa_{e,a} + \kappa_{i,a}}{2}\right) a - i\mu b - \sqrt{\kappa_{e,a}}\, a_{in}$$

$$\dot{b} = \left(-i\Delta_\mathrm{b} - \frac{\kappa_b}{2}\right) b - i\mu a$$

We have transitioned to the laser rotating frame $a \rightarrow a e^{i\omega_L t}, b \rightarrow b e^{i\omega_L t}$ with $\Delta_a = \omega_L - \omega_a$ and $\Delta_\mathrm{b} = \omega_L - \omega_b$ ($\omega_L$, $\omega_a$, $\omega_b$ are the frequency of laser, $a$, and $b$, respectively). Steady state solution ($a, b = 0$) gives an effective equation for $a$

$$0 = \left[-i(\Delta_a + \Delta_{eff}) - \frac{(\kappa_{e,a} + \kappa_{i,a} + \kappa_{eff})}{2}\right] a - \sqrt{\kappa_{e,a}}\, a_{in}$$

In which

$$\Delta_{eff} = -\Delta_\mathrm{b} \frac{\mu^2}{\frac{\kappa_b^2}{4} + \Delta_\mathrm{b}^2} \qquad \kappa_{eff} = \kappa_b \frac{\mu^2}{\frac{\kappa_b^2}{4} + \Delta_\mathrm{b}^2}$$

Therefore, coupling to a resonant mode $b$ introduces an effective detuning and loss rate for mode $a$. The GCC can be achieved when $\kappa_{e,a} = \kappa_{i,a} + \kappa_{eff}$ and $\Delta_a + \Delta_{eff} = 0$. By manipulating the loss rates under GCC, energy flow can be engineered. For instance, when $\kappa_{e,a} \sim \kappa_{eff} \gg \kappa_{i,a}$ , nearly 100% of the energy is directed towards mode $b$, thus preventing wastage of energy through channel $\kappa_{i,a}$.

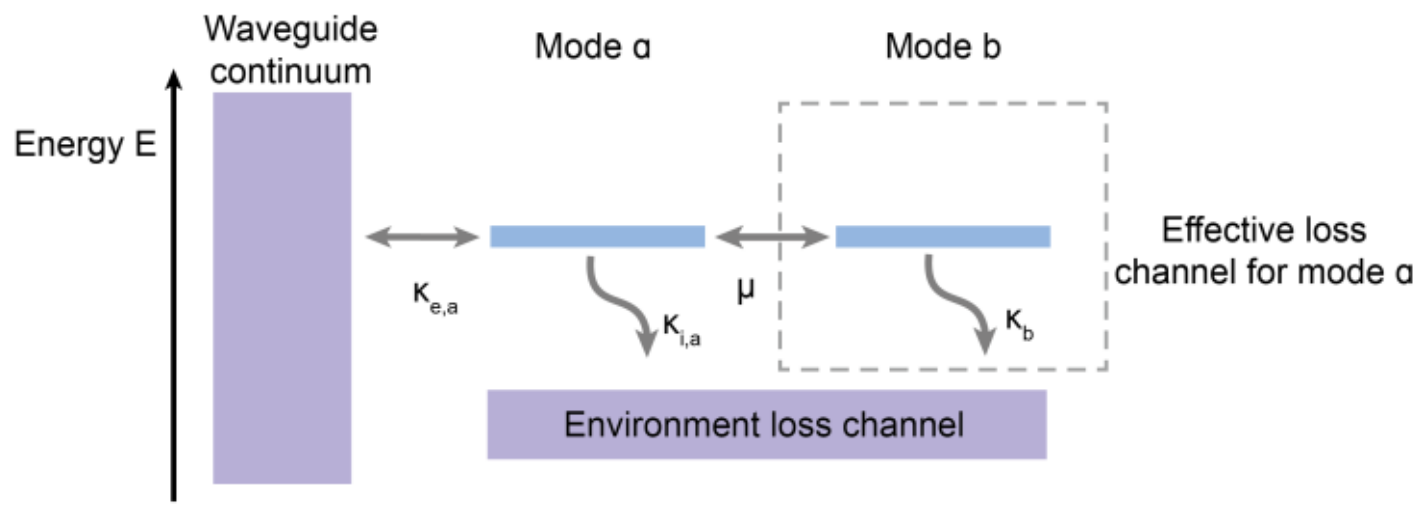

**Box Figure 2** | Illustration of the general critical coupling.

In many cases, it is common to simplify the system by setting $\Delta_\mathrm{a} = \Delta_b = 0$. In this case, the physical picture is: a resonance-coupling induces a loss channel equivalent to $\kappa_{eff} = 4\mu^2/\kappa_b$, with $\mu$ is the resonant coupling and $\kappa_b$ is the mode that causes this loss. This simplification aids in a quick estimation of the loss induced by resonance-coupling. Note that this also applies to multiple levels coupled as a chain. The effective loss created by a chain of levels can be obtained by cascading this equivalent loss calculation (e.g. the process of cavity EO comb generation[6]). The GCC can be a useful tool for manipulating energy flow in systems with multiple resonant modes, such as coupled-cavity system as well as applications in frequency shifting[3], EO comb[6], Kerr comb[225,226], nonlinear frequency conversion[227], and more.

**Box 3 Microwave engineering for EO modulation**

The bandwidth of an EO modulator critically relies on the microwave engineering of its electrode. Here we discuss some basic concepts of microwave engineering. The electrode is typically classified as lumped element or distributed element, determined by the electrode characteristic size $L$ and the wavelength of the microwave. For example, when the wavelength $\lambda_{MW}$ of the applied microwave is much larger than the electrode size $L$, the electrode can be treated as a pure capacitor with resistance and inductance. When $\lambda_{MW} \sim L$ or $\lambda_{MW} \ll L$, distributed models should be used.

In the simple case where the electrode is a capacitor, the modulator's frequency response is limited by the RC (resistance-capacitance) time-constant of the electrode. Long electrodes with narrow gaps can enhance interaction and reduce $V_\pi$, but they also increase capacitance, resulting in smaller bandwidth. To overcome this limit, traveling-wave modulators are required, where the electrode is designed to be a transmission line (distributed element), allowing microwave to co-propagate with the optical field at the same speed. In this case, velocity matching (microwave phase velocity matching optical group velocity), impedance matching, and microwave loss are important considerations. A detailed discussion of traveling-wave modulator design can be found in Section 3.2 of Ref [45]. Different from bulk LN modulator, where the dielectric environment of the transmission line is almost fixed, the microwave velocity (i.e., microwave index) in TFLN modulators can be effectively engineered by adjusting the thickness of buried and cladding oxide and achieve nearly perfect velocity matching.

To understand the microwave behavior of a practical EO modulator system, it is helpful to describe the RF source, cable, and modulator electrode in the form of a distributed circuit model (see Box Figure below). The RF cable and transmission line electrodes can both be modeled using $LC$ ladders, and the characteristic impedance is $\sqrt{L/C}$, where $L$ and $C$ are inductance and capacitance per unit length. Note that the source has an output impedance, $R_{\text{source}}$, commonly at 50 Ohm. When measuring low-frequency response of capacitive load or unterminated transmission line (a common configuration for $V_\pi$ measurement), the actual voltage on the device can be twice as the voltage displayed on the waveform/signal generator, since the displayed values are rated to matched load. This is a common mistake that can lead to underestimation of modulator Vpi in experimental measurement.

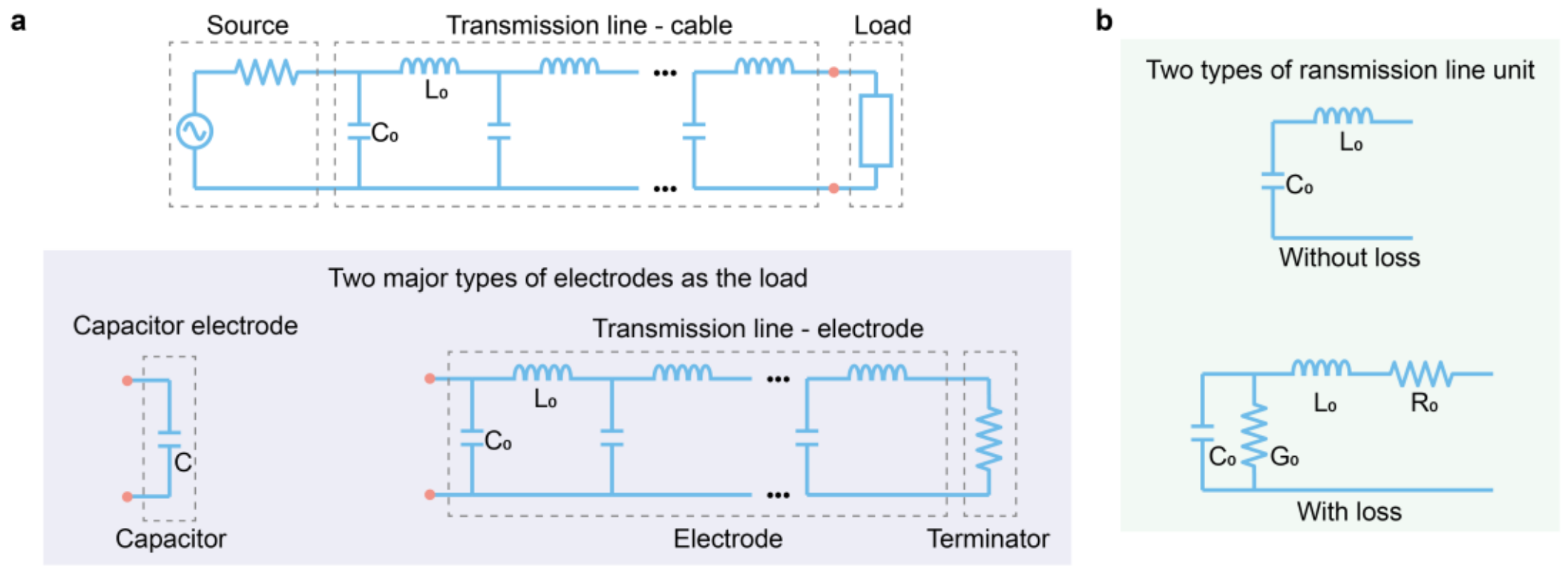

**Box Figure 3 | a,** Basic concept of microwave engineering for efficient modulation on TFLN. The transmission line impedance is given by $Z_0 = \sqrt{L_0/C_0}$. **b,** Illustration of transmission line unit cell with and without losses. In a), unit cells without loss are used as an example. The resistance $R_0$ and conductance $G_0$ contribute to the microwave propagation loss in the transmission line.

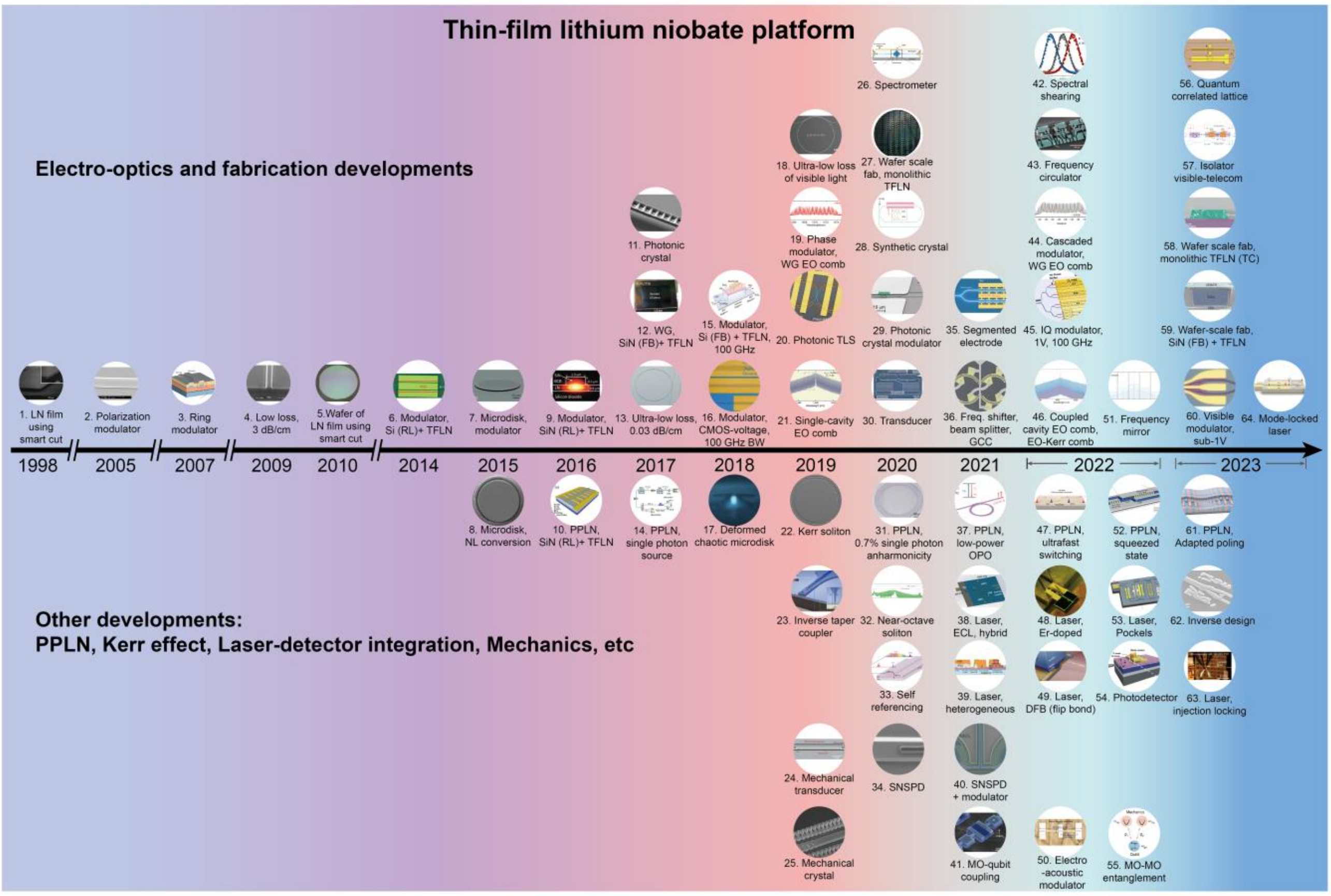

**Fig. 1 | Time-line of various thin-film lithium niobate platforms development.** Upper panel shows the development of TFLN electro-optics, while the bottom one shows the development of related TFLN platforms including periodically-poled TFLN (PP-TFLN), III-V-TFLN integration, etc. Type of integration is indicated for heterogeneous/hybrid integrated platforms, while those without the label are monolithic TFLN. **Abbreviations:** RL, rib-loaded; FB, film-bonding; NL, nonlinear; PPLN, periodically-poled lithium niobate; WG, waveguide; TLS, two-level-system; GCC, general critical coupling; MO, mechanical oscillator; SNSPD, superconducting nanowire single photon detector; OPO, optical parametric oscillation; ECL, external cavity laser; TC, tight-confining; DFB, distributed feedback; CMOS, complementary metal-oxide semiconductor; LN, lithium niobate; TFLN, thin-film lithium niobate; Si, silicon; SiN, silicon nitride; **References:** 1. LN film using smart cut: ref. [40], together with others[228]; 2. Polarization modulator: ref. [229]; 3. Ring modulator: ref. [95]; 4. Low loss, 3 dB/cm: ref. [41]; 5. Wafer of LN film using smart cut: ref. [230]; 6. Modulator, Si (RL) + TFLN: ref. [66], together with others[67–69]; 7. Microdisk, modulator: ref. [231], together with others[232–235]; 8. Microdisk, NL conversion: ref. [236], together with others[237–245]; 9. Modulator, SiN (RL) + TFLN: ref. [72], together with others[70,71,73–75]; 10. PPLN, SiN (RL) + TFLN: ref. [76]; 11. Photonic crystal: ref. [246], together with others[99,247–250];12. WG, SiN (FB) + TFLN: ref. [62], together with others[63]; 13. Ultra-low loss, 0.03 dB/cm: ref. [43]; 14. PPLN, single photon source: ref. [251], together with others[252]; 15. Modulator, Si (FB) + TFLN 100 GHz: ref. [60], together with others[61]; 116. Modulator, CMOS-voltage, 100 GHz BW: ref [1]. 17. Deformed chaotic microdisk: ref. [253]; 18. Ultra-low loss of visible light: ref. [44]; 19. Phase modulator, WG EO comb: ref. [48]; 20 Photonic TLS: ref. [8]; 21. Single cavity EO comb: ref. [2], together with others[123]; 22. Kerr soliton: ref. [167], together with others[166]; 23. Inverse taper coupler: ref. [254]; 24. Mechanical transducer: ref. [255], together with others[249]; 25. Mechanical crystal: ref. [248]; 26. Spectrometer: ref. [256]; 27. Wafer scale fab, monolithic TFLN: ref. [257]; 28. Synthetic crystal: ref. [14]; 29. Photonic crystal modulator: ref. [98]; 30. Transducer: ref. [19], together with others[18,20]; 31. PPLN, 0.7% single photon anharmonicity: ref. [176]; 32. Near-octave soliton: ref. [168]; 33. Self referencing: ref. [258]; 34. SNSPD: ref. [259], together with others[260]; 35. Segmented electrode: ref. [52], together with others[82–84]; 36.

Freq. shifter, beam splitter, GCC: ref. [3]; 37. PPLN low-power OPO: ref. [261], together with others[262]; 38 Laser, ECL, hybrid: ref. [157]; 39. Laser, heterogeneous: ref. [163], together with others[162,164,165]; 40. SNSPD + modulator: ref. [263]; 41. MO-qubit coupling: ref. [264]; 42. Frequency circulator: ref. [7]; 43. Spectral shearing: ref. [12]; 44. Cascaded modulator, WG EO comb: ref. [9], together with others[111]; 45. IQ modulator: ref. [5], together with others[85,86]; 46. Coupled cavity EO comb, EO-Kerr comb: ref. [6]; 47. PPLN, ultrafast switching: ref. [265]; 48. Laser, Er-doped: ref. [156], together with others[155]; 49. Laser, DFB (flip bond): ref. [160]; 50. Electro-acoustic modulator: ref. [195]; 51. Frequency mirror: ref. [13]; 52. PPLN, squeezed state: ref. [193]; 53. Laser, Pockels: ref. [159]; 54. Photodetector: ref. [169], together with others[164,170]; 55. MO-MO entanglement: ref. [266]; 56. Quantum correlated lattice: ref. [15]; 57. Visible-telecom isolator: ref. [267], together with others[268]; 58. Wafer scale fab, monolithic (TC): ref. [269]; 59. Wafer scale fab, SiN (FB) + TFLN: ref. [64], together with others[65,158]; 60. Visible modulator, sub-1V: ref. [11], together with others[10]; 61. PPLN, adapted poling: ref. [270]; 62. Inverse design: ref. [271]; 63. Laser, injection locking: ref. [158]; 64. Mode-locked laser: ref. [220];

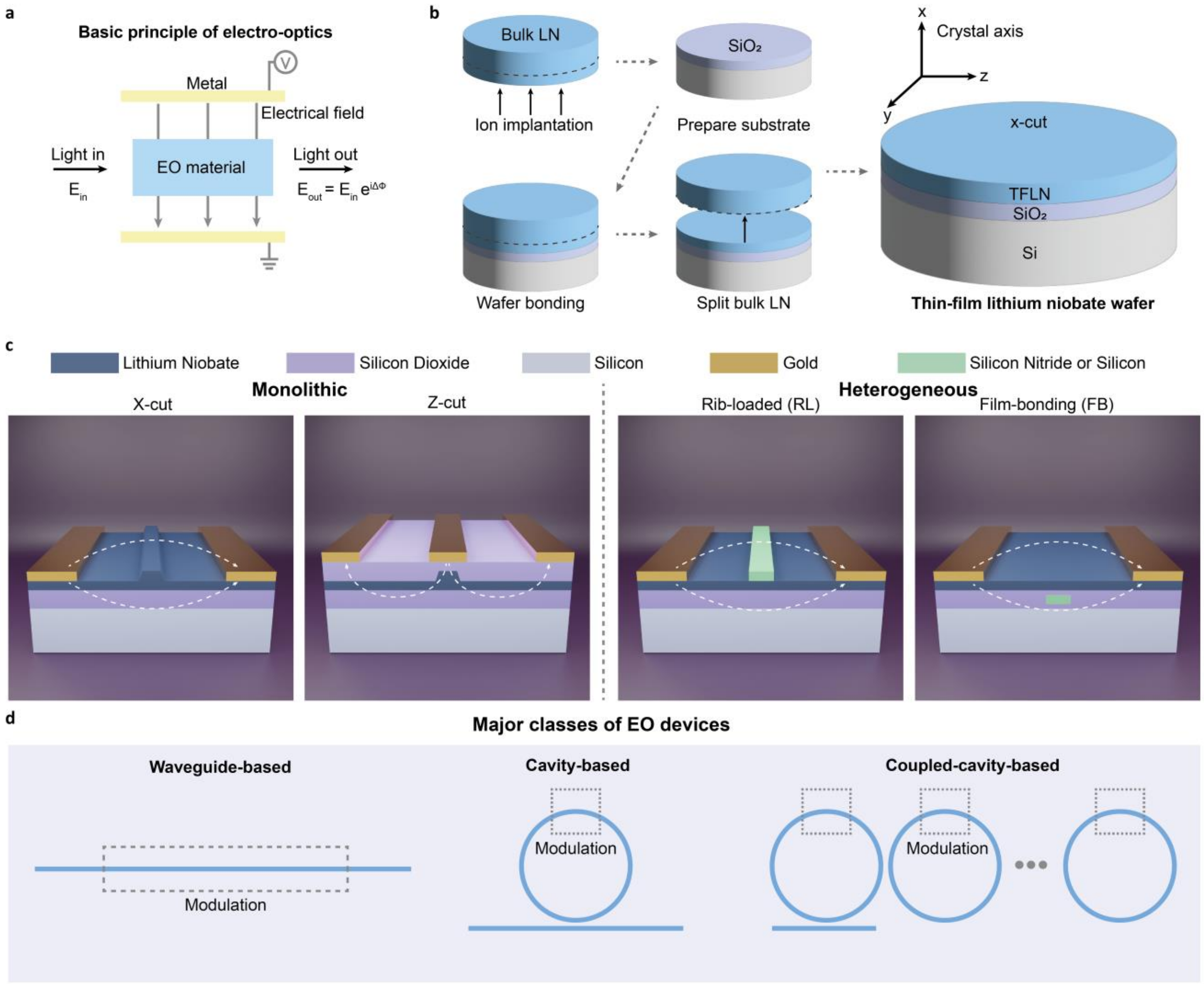

**Fig. 2 | Basic principle and device structures on thin-film lithium niobate. a,** Light sent into an EO material will acquire a phase shift that is proportional to the voltage applied to the material. **b,** Illustration of TFLN wafer and the "smart cut" technology used to prepare the TFLN wafer. The axis represents the crystal axis of the LN crystal. **c,** Cross-section for major types of TFLN EO platforms. The white dashed arrows denote the microwave electric field line. **d,** Illustration of three major classes of EO devices: waveguide, cavity, and coupled-cavity. The dashed box is used to label the electrode region for applying EO modulation.

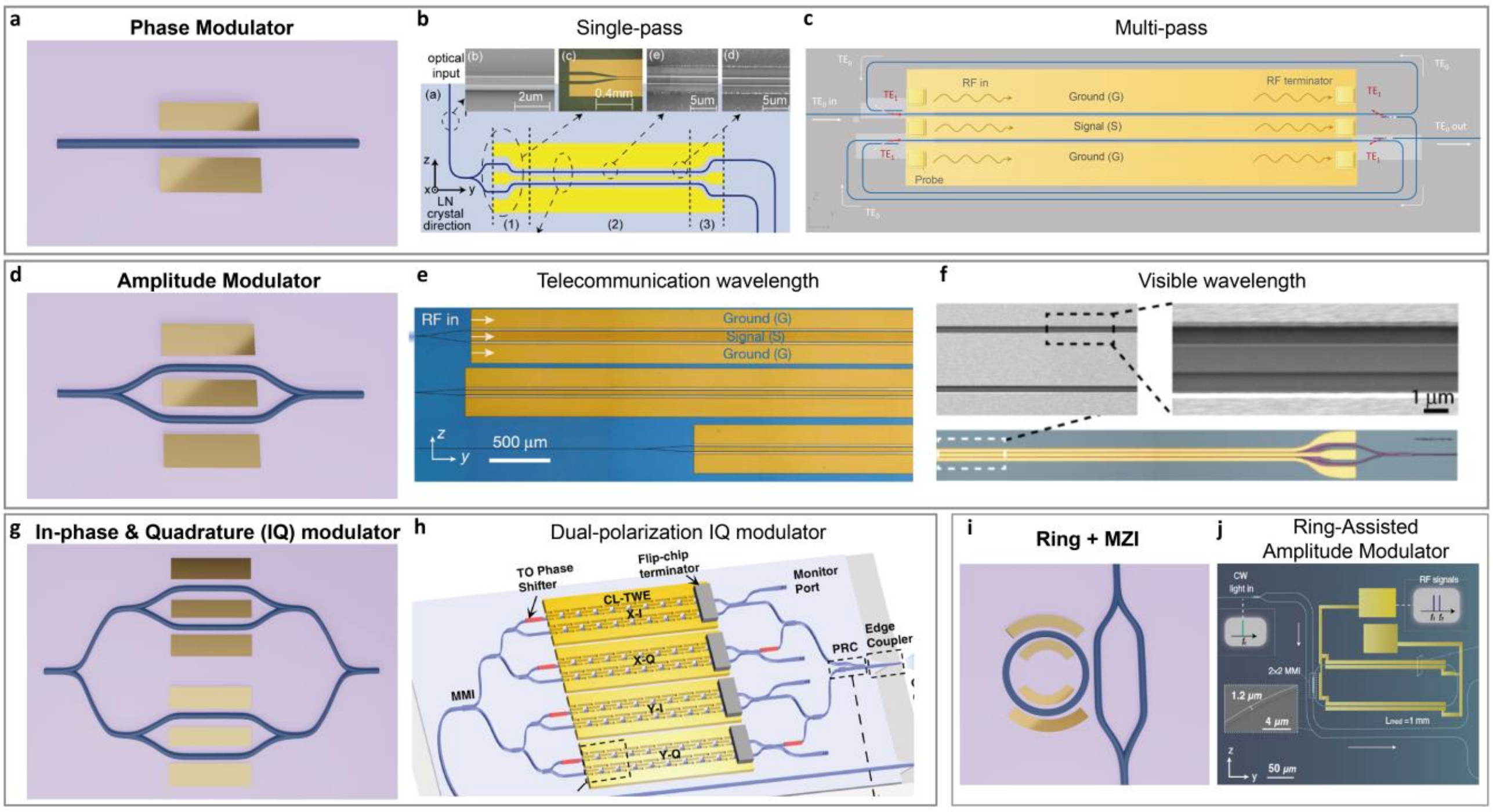

**Fig. 3 | Waveguide-based devices on thin-film lithium niobate. a,** Schematic configuration of PM. It contains a single waveguide with electrode. Voltage changes the phase of the light. **b,** Representative device image of a conventional TFLN PM. Input light is split into two paths. Light transmitted through the electrode region obtains a phase modulation. Two output ports are designed to receive light from each path. **c,** Representative device image of multi-pass PM. Light can transmit through the phase modulation region four-times through TE0/TE1 mode conversion. **d,** Schematic configuration of AM. Light is split into two paths with reversed-sign phase modulation. Interference occurs after passing through the electrode via the output splitter, converting the phase modulation to an amplitude modulation. **e,** Representative device image of AM with CMOS-compatible voltage and 100 GHz EO bandwidth at telecommunication wavelength. **f,** Representative device image of an AM at visible wavelength, featuring a $V_\pi \cdot L$ as low as 0.17 V $\cdot$ cm at 400nm. **g,** Schematic configuration of an IQ modulator. Two amplitude modulators form an interferometer, with each paths having a $\frac{\pi}{2}$ relative phase delay on microwave drive signal, allowing modulation in the complex space of in-phase and quadrature components. **h,** Representative device image of a dual-polarization IQ modulator, in which two IQ modulators with different polarizations are used in parallel and combined in the end, allowing polarization multiplexing. **i,** Schematic configuration of a ring-assisted AM. **j,** Representative device image of a ring-assisted AM for improving the linearity of an AM. Panels adapted with permission from: **b,** ref [48], IEEE; **c,** ref [91], Springer Nature Ltd; **e,** ref [1], Springer Nature Ltd; **f,** ref [11], Optica; **h,** ref [5], Optica; **j,** ref [90], Springer Nature Ltd;

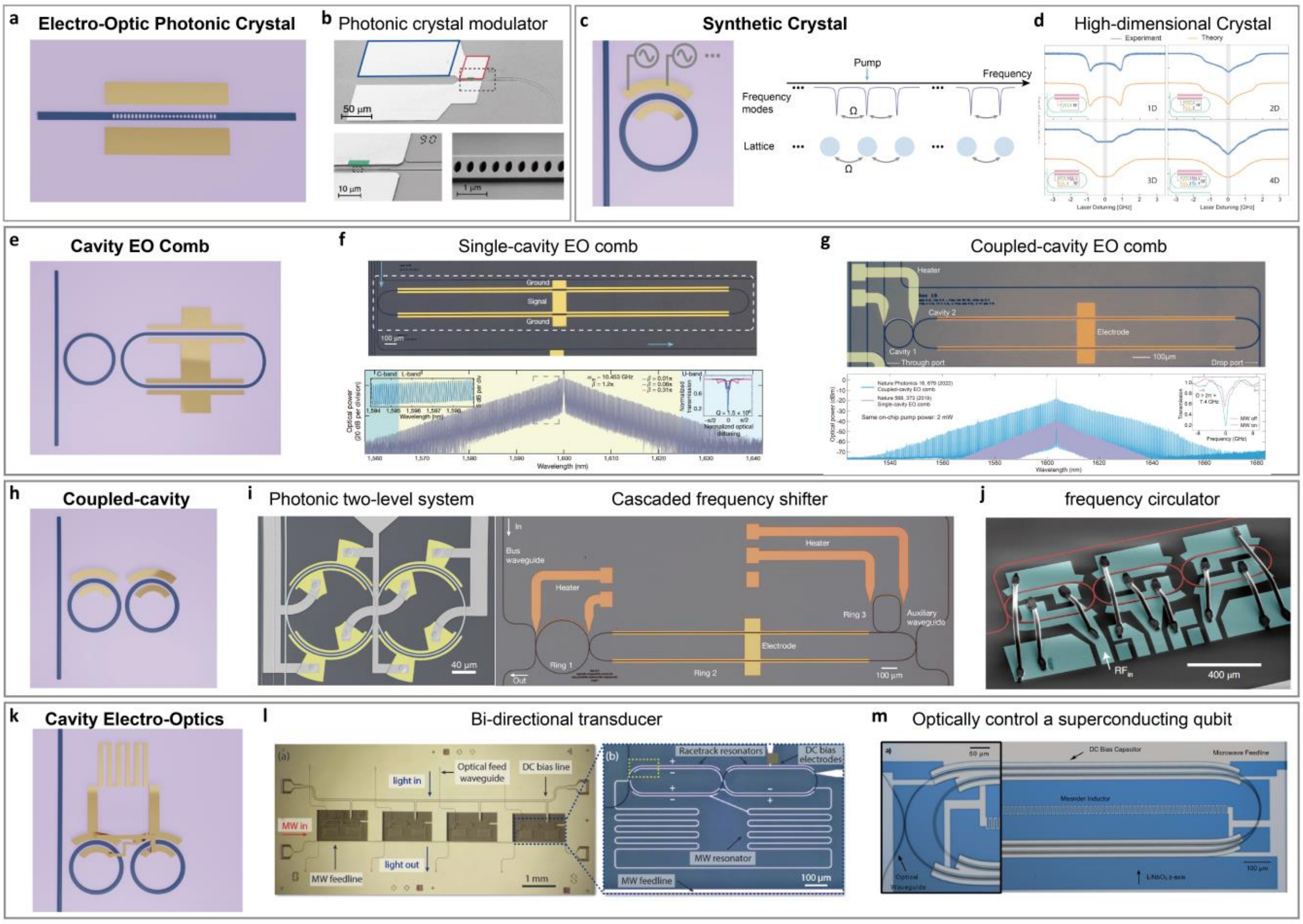

**Fig. 4 | Cavity-based EO devices on thin-film lithium niobate. a,** Schematic configuration of a photonic crystal modulator. **b,** Representative device image of photonic crystal modulator. **c,** Schematic of a synthetic EO crystal. **d,** Representative data of a high-dimensional frequency crystal. **e,** schematic configuration of a coupled-cavity EO comb. **f,** representative device image of a single cavity EO comb; **g,** representative device image of a coupled-cavity EO comb; **h,** schematic configuration of a double-ring coupled-cavity EO modulator. Two identical rings form a photonic TLS. Microwave modulation drives the transition. **i,** representative device image of a photonic TLS as well as a frequency shifter and beam splitter. In a cascaded frequency shifter, light flows through a ladder of energy levels. This allows one to shift the light frequency beyond 100 GHz using only a continuous and single-tone microwave at several tens of GHz. **j,** Representative device image of a triple-ring coupled-cavity frequency circulator, enabling 40 dB isolation with only 75 mW microwave power. **k,** Schematic configuration of a cavity electro-optics device, formed by a lumped-element microwave cavity coupled to an optical cavity. **l, m,** representative device image for cavity electro-optics, including transducers and optical control of a superconducting qubit. Panels adapted with permission from: **b,** ref. [98], Springer Nature Ltd; **d,** ref. [14], Optica; **f,** ref. [2] Springer Nature Ltd; **g,** ref. [6], Springer Nature Ltd; **i,** ref [3], Springer Nature Ltd; **j,** ref. [7], Springer Nature Ltd; **l,** ref. [19], Optica; **m,** ref.[145], arXiv;

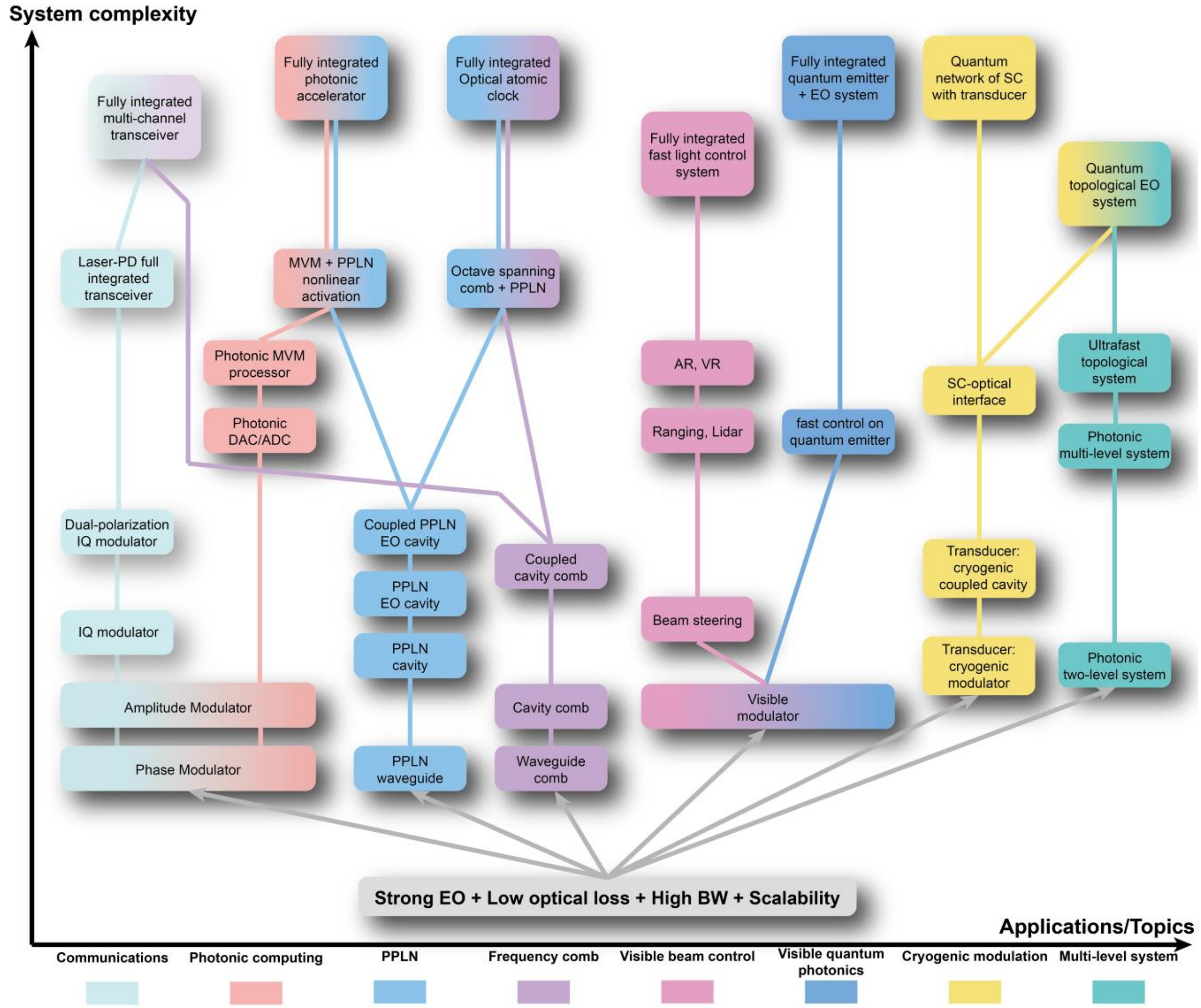

**Fig. 5 | Evolution and outlook for applications of integrated EO system on thin-film lithium niobate.** The height of the tree denotes the system complexity while each branch of the tree represents one topic or application. Some of the branches merge as the complexity increases, leading to more advanced systems. Abbreviation in the figure: MVM, matrix-vector multiplication; DAC, digital-to-analog converter; ADC, analog-to-digital converter; PD, photodiode; PPLN, periodically poled lithium niobate; IQ, in-phase and quadrature; AM, amplitude modulator; PM, phase modulator; EO, electro-optics; AR, augmented reality; VR, virtual reality; SC, superconducting qubit.